\documentclass[12pt]{article}
\usepackage{amssymb}

\begin{document}

\bigskip\ 

\bigskip\ 

\begin{center}
\textbf{TOWARDS A BACKGROUND INDEPENDENT}

\smallskip\ 

\textbf{QUANTUM GRAVITY IN EIGHT DIMENSIONS}

\textbf{\ }

\textbf{\ }

\smallskip\ 

J. A. Nieto\footnote{%
nieto@uas.uasnet.mx}

\smallskip

\textit{Facultad de Ciencias F\'{\i}sico-Matem\'{a}ticas de la Universidad
Aut\'{o}noma}

\textit{de Sinaloa, 80010, Culiac\'{a}n Sinaloa, M\'{e}xico}

\bigskip\ 

\bigskip\ 

\textbf{Abstract}
\end{center}

We start a program of background independent quantum gravity in eight
dimensions. We first consider canonical gravity \textit{a la} "Kaluza-Klein"
in $D=d+1$ dimensions. We show that our canonical gravity approach can be
applied to the case of self-dual gravity in four dimensions. Further, by
using our previously proposed classical action of Ashtekar self-dual gravity
formalism in eight dimensions, we proceed to develop the canonical approach
in eight dimensions. Our construction considers different $SO(8)$ symmetry
breakings. In particular, the breaking $SO(8)=S_{R}^{7}\times
S_{L}^{7}\times G_{2}$ plays an important role in our discussion.

\bigskip\ 

\bigskip\ 

\bigskip\ 

\bigskip\ 

\bigskip\ 

Keywords: Ashtekar theory, eight dimensions, octonions.

Pacs numbers: 04.60.-m, 04.65.+e, 11.15.-q, 11.30.Ly

October, 2008

\newpage

\bigskip\ 

\noindent \textbf{1. Introduction}

\smallskip\ 

Considering the motivation for background independent quantum gravity [1]
one finds that most of the arguments can be applied not only to four
dimensions but to any higher dimensional gravitational theory based in
Einstein-Hilbert action. For instance, the statement that "gravity is
geometry and therefore there should no be background metric" is also true in
a higher dimensional gravitational theory based in Einstein-Hilbert action.
Similar conclusion can be obtained thinking in a non-perturbative context.
So, why to rely only in four dimensions when one considers background
independent quantum gravity? Experimental evidence of general relativity in
four dimensions is established only at the classical, but not at the quantum
level. Thus at present, in the lack of experimental evidence of quantum
gravity any argument concerning the dimensionality of the spacetime should
be theoretical.

A possibility for setting four dimensions comes from the proposal of
self-dual gravity [2]-[3]. One starts with the observation that the
potential (playing by the three dimensional scalar curvature) in the
Hamiltonian constraint is difficult to quantize. In the case of four
dimensions it is shown that such a potential can be avoided by introducing
new canonical variables [4] which eventually are obtained via self-dual
gravity [2]-[3]. In turn, self-dual gravity seems to make sense only in four
dimensions since in this case the dual of a two form (the curvature) is
again a two form. This argument is based on the definition of the duality
concept in terms of the completely antisymmetric density $\epsilon
_{A_{0}..A_{D-1}}$ which takes values in the set $\{-1,0,1\}$. The Riemann
curvature $R^{AB}$ is a two form. Thus the dual $^{\ast }R_{A_{0}...A_{D-3}}=%
\frac{1}{2}\epsilon _{A_{0}..A_{D-3}A_{D-2}A_{D-1}}R^{A_{D-2}A_{D-1}}$ is a
two form only for $D=4$. Hence, in trying to define the self-dual object $%
^{+}R^{AB}$ one discovers that only in four dimensions one can establish the
combination $^{+}R^{AB}=\frac{1}{2}(R^{AB}-i^{\ast }R^{AB})$.

The definition of duality in terms of the $\epsilon $-symbol is not,
however, the only possibility. A number of authors [5]-[8] have shown that
duality also makes sense through a definition in terms of the $\eta $%
-symbol. In fact, the $\eta $-symbol is very similar to the $\epsilon $%
-symbol in four dimensions; is a four index completely antisymmetric object
and take values also in the set $\{-1,0,1\}$. However, the $\eta $-symbol
lives in eight dimensions rather than in four. Moreover, while the $\epsilon 
$-symbol in four dimensions can be connected with quaternions, the $\eta $%
-symbol is related to the structure constants of octonions (see [9]-[10] and
Refs. therein). Thus, in eight dimensions we can also introduce the dual $%
^{\star }R_{A_{0}A_{1}}=\frac{1}{2}\eta
_{A_{0}A_{1}A_{2}A_{3}}R^{A_{2}A_{3}} $ and consequently the self-dual
object $^{+}R^{AB}=\frac{1}{4}(R^{AB}+^{\star }R^{AB})$ (see section 6 for
details). It remains to prove whether by using this new kind of duality we
can also avoid the potential in terms of the scalar Riemann curvature in the
Hamiltonian constraint which is inherent to any higher dimensional theory as
we shall see in section 2. In this work we show that in fact duality in
terms of the $\eta $-symbol avoids also such a potential. Our strategy is
first to develop canonical gravity \textit{a la }"Kaluza-Klein" and then to
discuss self-dual gravity in four dimensions. This allows us to follow a
parallel program in eight dimensions and in this way to determine the
canonical constraints of self-duality gravity in eight dimensions.

The above comments can be clarified further with the help of group theory.
We recall that in four dimensions the algebra $so(1,3)$ can be written as $%
so(1,3)=su(2)\times su(2)$. So, the curvature $R^{AB}$ can be decomposed
additively [2]: $R^{AB}(\omega )=\quad ^{+}R^{AB}(^{+}\omega
)+^{-}R^{AB}(^{-}\omega )$ where $^{+}\omega $ and $^{-}\omega $ are the
self-dual and anti-self-dual parts of the spin connection $\omega $. In an
Euclidean context this is equivalent to write the norm group of quaternions $%
O(4)$ as $O(4)=S^{3}\times S^{3}$, where $S^{3}$ denotes the three sphere.
The situation in eight dimensions is very similar since $O(8)=S^{7}\times
S^{7}\times G_{2}$, with $S^{7}$ denoting the seven sphere, suggesting that
one can also define duality in eight dimensions, but modulo the exceptional
group $G_{2}$ [11]-[12].

In turn, these results in the context of group theory are connected with the
famous Hurwitz theorem which establishes that any normed algebra is
isomorphic to the following: real, complex, quaternion and octonion algebra
(see [10] and Refs. therein). Considering duality, one learns that it is
reasonable to define it for quaternions and octonions via the generalized
vector product [11]. In this sense, the classical approach of Ashtekar
formalism in eight dimensions proposed in Refs. [13]-[15] has some kind of
uniqueness. In this work we give some steps forward on the program of
developing quantum gravity in eight dimensions. Specifically, in sections 6,
by using self-dual gravity defined in terms of the $\eta $-symbol we develop
a canonical gravity in eight dimensions. We find the eight dimensional
canonical Diffeomorphism and Hamiltonian constraints and we outline, in the
final section, a possible physical quantum states associated with such
constraints.

\bigskip\ 

\noindent \textbf{2.\ Canonical gravity }$a$ $la$\textbf{\ "Kaluza-Klein"}

\smallskip\ 

Let us start with a brief review of canonical gravity. We shall use some
kind of "Kaluza-Klein" mechanism for our review. One of the advantage of
this method is that one avoids the use of a time-like vector field. This
allows us to describe, in straightforward way, canonical self-dual gravity
at the level of the action for both four and eight dimensions. Although our
canonical method resembles the one used in Ref. [16] our approach contains
complementary descriptions and computations.

We shall assume that the vielbein field $e_{\mu }^{~(A)}=e_{\mu
}^{~(A)}(t,x),$ on a $D=d+1$-manifold $M^{D}$, can be written in the form

\begin{equation}
e_{\mu }^{~(A)}=\left( 
\begin{array}{cc}
E_{0}^{~(0)}(t,x) & E_{0}^{~(a)}(t,x) \\ 
0 & E_{i}^{~(a)}(t,x)%
\end{array}%
\right) .  \label{1}
\end{equation}%
Although in writing (1) we do not consider any kind of dimensional reduction
or compactification, this form of $e_{\mu }^{~(A)}$ is in a sense inspired
by the Kaluza-Klein mechanism. The inverse $e_{(A)}^{~~~\mu }$ can be
obtained from the relation $e_{\nu }^{~(A)}e_{(A)}^{~~~\mu }=\delta _{\nu
}^{\mu }$, with $\delta _{\nu }^{\mu }$ denoting the Kronecker delta. We find

\begin{equation}
e_{(A)}^{~~~\mu }=\left( 
\begin{array}{cc}
E_{(0)}^{~~~0}(t,x) & E_{(0)}^{~~~i}(t,x) \\ 
0 & E_{(a)}^{~~~i}(t,x)%
\end{array}%
\right) ,  \label{2}
\end{equation}%
with $%
E_{(0)}^{~~~0}=1/E_{0}^{~(0)},E_{(0)}^{~~~i}=-E_{0}^{~(a)}E_{(a)}^{~~~i}/E_{0}^{~(0)} 
$ and $E_{j}^{~(a)}E_{(a)}^{~~~i}=\delta _{j}^{i}$. In the above the indices 
$(A)$ and $\mu $ of $e_{\mu }^{~(A)}$ denote frame and target spacetime
indices respectively.

In general, the metric $\gamma _{\mu \nu }$ is defined in terms of $e_{\mu
}^{~(A)}$ in the usual form

\begin{equation}
\gamma _{\mu \nu }=e_{\mu }^{~(A)}e_{\nu }^{~(B)}\eta _{(AB)}.  \label{3}
\end{equation}%
Here, $\eta _{(AB)}$ is a flat $(d+1)$-metric. We shall write $e_{\mu
(A)}=e_{\mu }^{~(B)}\eta _{(AB)}$, $e^{(A)\mu }=e_{(B)}^{~~~\mu }\eta
^{(AB)} $ and also $e_{\mu (A)}=\gamma _{\mu \nu }e_{(A)}^{~~~\nu }$ and $%
e^{(A)\mu }=\gamma ^{\mu \nu }e_{\nu }^{~(A)}$, where $\eta ^{(AB)}$ is the
inverse of $\eta _{(AB)}$.

In the particular case in which $e_{\mu }^{~(A)}$ is written as (1) $\gamma
_{\mu \nu }$ becomes

\begin{equation}
\gamma _{\mu \nu }=\left( 
\begin{array}{cc}
-N^{2}+g_{ij}N^{i}N^{j} & N_{i} \\ 
N_{j} & g_{ij}%
\end{array}%
\right) ,  \label{4}
\end{equation}%
where $N=E_{0}^{~(0)}$, $N_{i}=E_{0}^{~(a)}E_{i}^{~(b)}\delta _{(ab)}$, $%
g_{ij}=E_{i}^{~(a)}E_{j}^{~(b)}\delta _{(ab)}$ and $N^{i}=g^{ij}N_{j}$, with 
$g^{ik}g_{kj}=\delta _{j}^{i}.$ Here the symbol $\delta _{(ab)}$ also
denotes a Kronecker delta.

We also find that

\begin{equation}
\gamma ^{\mu \nu }=\left( 
\begin{array}{cc}
-N^{-2} & N^{-2}N^{i} \\ 
N^{-2}N^{j} & g^{ij}-N^{-2}N^{i}N^{j}%
\end{array}%
\right) .  \label{5}
\end{equation}%
We observe that (4) and (5) provide the traditional ansatz for canonical
gravity. So, $N$ and $N_{i}$ admit the interpretation of lapse function and
shift vector, respectively. Thus, in terms of $N$ and $N_{i}$, (1) and (2)
become

\begin{equation}
e_{\mu }^{~(A)}=\left( 
\begin{array}{cc}
N & E_{i}^{~(a)}N^{i} \\ 
0 & E_{i}^{~(a)}%
\end{array}%
\right)  \label{6}
\end{equation}%
and%
\begin{equation}
e_{(A)}^{~~~\mu }=\left( 
\begin{array}{cc}
N^{-1} & -N^{-1}N^{i} \\ 
0 & E_{(a)}^{~~~i}%
\end{array}%
\right) .  \label{7}
\end{equation}%
For later calculations it is convenient to write $E_{i(a)}=E_{i}^{~(a)}\eta
_{(ab)}$, $E^{(a)i}=E_{(b)}^{~~~i}\eta ^{(ab)}$ and also $%
E_{i(a)}=g_{ij}E_{(a)}^{~~~j}$, $E^{(a)i}=g^{ij}E_{j}^{~(a)}.$ Observe that
although $e_{i}^{~(a)}=E_{i}^{~(a)}$ we have $e^{(a)i}\neq E^{(a)i}$. This
is because when we consider the $e$ notation we raise and lower indices with
the metric $\gamma $, while in the case of the $E$ notation we raise and
lower indices with the metric $g.$ In fact, this is one of the reasons for
distinguishing $e$ and $E$ in the ansatz (1) and (2).

We shall assume that $e_{\mu }^{~(A)}$ satisfies the condition

\begin{equation}
\partial _{\mu }e_{\nu }^{~(A)}-\Gamma _{\mu \nu }^{\alpha }e_{\alpha
}^{~(A)}+\omega _{\mu }^{~(AB)}e_{\nu (B)}=0.  \label{8}
\end{equation}%
Here, $\Gamma _{\mu \nu }^{\alpha }(\gamma )=\Gamma _{\nu \mu }^{\alpha
}(\gamma )$ and $\omega _{\nu }^{~(AB)}=-\omega _{\nu }^{~(BA)}$ denote the
Christoffel symbols and the spin connection respectively. The expression (8)
determines, of course, a manifold with a vanishing torsion. Using (8), it is
not difficult to see that $\omega _{(ABC)}=e_{(A)}^{~~~\mu }\omega _{\mu
(BC)}=-$ $\omega _{(ACB)}$ can be written in terms of%
\begin{equation}
F_{\mu \nu }^{~~~(A)}=\partial _{\mu }e_{\nu }^{~(A)}-\partial _{\nu }e_{\mu
}^{~(A)}  \label{9}
\end{equation}%
in the following form

\begin{equation}
\omega _{(ABC)}=\frac{1}{2}\left[ F_{(ABC)}+F_{(CAB)}+F_{(CBA)}\right] ,
\label{10}
\end{equation}%
where%
\begin{equation}
F_{(ABC)}=e_{(A)}^{~~~\mu }e_{(B)}^{~~~\nu }F_{\mu \nu (C)}=-F_{(BAC)}.
\label{11}
\end{equation}

Considering (6), (7) and (9) we find

\begin{equation}
F_{0i(0)}=\partial _{i}N,  \label{12}
\end{equation}

\begin{equation}
F_{ij(0)}=0,  \label{13}
\end{equation}%
\begin{equation}
F_{0i(a)}=\partial _{0}E_{i(a)}-\partial _{i}E_{j(a)}N^{j}-E_{j(a)}\partial
_{i}N^{j}  \label{14}
\end{equation}%
and

\begin{equation}
F_{ij(a)}=\partial _{i}E_{j(a)}-\partial _{j}E_{i(a)}.  \label{15}
\end{equation}%
Our aim is to obtain the different components of $\omega _{\mu (BC)}$
knowing the expressions (12)-(15). For this purpose we first observe that
(13) implies

\begin{equation}
F_{(ab0)}=0.  \label{16}
\end{equation}%
Thus, (10) leads to the following splitting%
\begin{equation}
\omega _{(00a)}=F_{(a00)},  \label{17}
\end{equation}

\begin{equation}
\omega _{(0ab)}=\frac{1}{2}\left[ F_{(0ab)}-F_{(0ba)}\right] ,  \label{18}
\end{equation}%
\begin{equation}
\omega _{(a0b)}=\frac{1}{2}\left[ F_{(a0b)}+F_{(b0a)}\right] ,  \label{19}
\end{equation}%
and%
\begin{equation}
\omega _{(abc)}=\frac{1}{2}\left[ F_{(abc)}+F_{(cab)}+F_{(cba)}\right] .
\label{20}
\end{equation}%
Since

\begin{equation}
\omega _{i(0a)}=E_{i}^{~(b)}\omega _{(b0a)},  \label{21}
\end{equation}

\begin{equation}
\omega _{0(bc)}=N\omega _{(0bc)}+E_{i}^{~(a)}N^{i}\omega _{(abc)},
\label{22}
\end{equation}%
\begin{equation}
\omega _{0(0b)}=N\omega _{(00b)}+E_{i}^{~(a)}N^{i}\omega _{(a0b)},
\label{23}
\end{equation}%
and%
\begin{equation}
\omega _{i(abc)}=E_{i}^{~(a)}\omega _{(abc)},  \label{24}
\end{equation}%
by means of (6)-(7) we get

\begin{equation}
\begin{array}{c}
\omega _{i(0a)}=\frac{N^{-1}}{2}%
E_{i}^{~(b)}[E_{(b)}^{~~~j}F_{j0(a)}-E_{(b)}^{~~~j}N^{k}F_{jk(a)} \\ 
\\ 
+E_{(a)}^{~~~j}F_{j0(b)}-E_{(a)}^{~~~j}N^{k}F_{jk(b)}],%
\end{array}
\label{25}
\end{equation}

\begin{equation}
\begin{array}{c}
\omega _{0(bc)}=\frac{N^{-1}}{2}\left[
E_{(b)}^{~~~i}F_{0i(c)}-N^{i}E_{(b)}^{~~~j}F_{ij(c)}-E_{(c)}^{~~~i}F_{0i(b)}+N^{i}E_{(c)}^{~~~j}F_{ij(b)}%
\right] \\ 
\\ 
+E_{i}^{~(a)}N^{i}\omega _{(abc)},%
\end{array}
\label{26}
\end{equation}%
and%
\begin{equation}
\begin{array}{c}
\omega _{0(0b)}=NF_{(b00)}+\frac{N^{-1}}{2}%
E_{k}^{~(a)}N^{k}[E_{(a)}^{~~~i}F_{i0(b)}-E_{(a)}^{~~i}N^{j}F_{ij(b)} \\ 
\\ 
+E_{(b)}^{~~~i}F_{i0(a)}-E_{(b)}^{~~~i}N^{j}F_{ij(a)}].%
\end{array}
\label{27}
\end{equation}%
Consequently, using (12)-(15) it is not difficult to obtain the results%
\begin{equation}
\omega _{i(0a)}=\frac{N^{-1}}{2}E_{(a)}^{~~~j}\left[ -\partial
_{0}g_{ij}+D_{i}N_{j}+D_{j}N_{i}\right] ,  \label{28}
\end{equation}

\begin{equation}
\begin{array}{c}
\omega _{0(bc)}=\frac{N^{-1}}{2}[E_{(b)}^{~~~i}\partial
_{0}E_{i(c)}-E_{(c)}^{~~~i}\partial _{0}E_{i(b)} \\ 
\\ 
-(E_{(b)}^{~~~i}E_{(c)}^{~~~j}-E_{(c)}^{~~~i}E_{(b)}^{~~~j})D_{i}N_{j}]%
\end{array}
\label{29}
\end{equation}%
and 
\begin{equation}
\omega _{0(0b)}=-E_{(b)}^{~~~i}\partial _{i}N+\frac{N^{-1}}{2}%
N^{i}E_{(b)}^{~~~j}\left[ -\partial _{0}g_{ij}+D_{i}N_{j}+D_{j}N_{i}\right] ,
\label{30}
\end{equation}%
where $D_{i}$ denotes covariant derivative in terms of the Christoffel
symbols $\Gamma _{jk}^{i}=\Gamma _{jk}^{i}(g)$.

With the help of (28), (29) and (30), we are now ready to compute the
Riemann tensor

\begin{equation}
R_{\mu \nu (AB)}=\partial _{\mu }\omega _{\nu (AB)}-\partial _{\nu }\omega
_{\mu (AB)}+\omega _{\mu (AC)}\omega _{\nu ~~(B)}^{~(C)}-\omega _{\nu
(AC)}\omega _{\mu ~~(B)}^{~(C)}.  \label{31}
\end{equation}%
But before we do that let us first observe that

\begin{equation}
R_{ij(0a)}=\mathcal{D}_{i}\omega _{j(0a)}-\mathcal{D}_{j}\omega _{i(0a)},
\label{32}
\end{equation}%
where

\begin{equation}
\mathcal{D}_{i}\omega _{j(0a)}=\partial _{i}\omega _{j(0a)}-\Gamma
_{ij}^{k}(g)\omega _{j(0a)}-\omega _{j(0c)}\omega _{i~~(a)}^{~(c)}.
\label{33}
\end{equation}%
We also obtain

\begin{equation}
R_{ij(ab)}=\tilde{R}_{ij(ab)}+\omega _{i(0a)}\omega _{j(0b)}-\omega
_{j(0a)}\omega _{i(0b)},  \label{34}
\end{equation}%
\begin{equation}
R_{0i(0a)}=\partial _{0}\omega _{i(0a)}-\partial _{i}\omega _{0(0a)}+\omega
_{0(0c)}\omega _{i~~(a)}^{~(c)}-\omega _{i(0c)}\omega _{0~~(a)}^{~(c)}
\label{35}
\end{equation}%
and%
\begin{equation}
\begin{array}{c}
R_{0i(ab)}=\partial _{0}\omega _{i(ab)}-\partial _{i}\omega _{0(ab)}+\omega
_{0(ac)}\omega _{i~~(b)}^{~(c)}-\omega _{i(ac)}\omega
_{0~~(b)}^{~(c)}+\omega _{0(0a)}\omega _{i(0b)} \\ 
\\ 
-\omega _{i(0a)}\omega _{0(0b)}.%
\end{array}
\label{36}
\end{equation}%
Here,

\begin{equation}
\tilde{R}_{ij(ab)}=\partial _{i}\omega _{j(ab)}-\partial _{i}\omega
_{j(ab)}+\omega _{i(ac)}\omega _{j~~(b)}^{~(c)}-\omega _{j(ac)}\omega _{\mu
~~(b)}^{~(c)}.  \label{37}
\end{equation}%
It becomes convenient to write

\begin{equation}
K_{ij}=\frac{N^{-1}}{2}\left( -\partial
_{0}g_{ij}+D_{i}N_{j}+D_{j}N_{i}\right) .  \label{38}
\end{equation}%
So, by using (28)-(30) we get

\begin{equation}
R_{ij(ab)}=\tilde{R}_{ij(ab)}+\left[
E_{(a)}^{~~~k}E_{(b)}^{~~~l}K_{ik}K_{jl}-E_{(a)}^{k}E_{(b)}^{l}K_{jk}K_{il}%
\right] ,  \label{39}
\end{equation}%
\begin{equation}
\begin{array}{c}
R_{0i(0a)}=\partial _{0}(E_{(a)}^{~~~k})K_{ik}+E_{(a)}^{~~~k}\partial
_{0}K_{ik}-\frac{1}{2}E^{(c)k}K_{ik}[E_{(c)}^{~~~l}\partial _{0}E_{l(a)} \\ 
\\ 
-E_{(a)}^{~~~l}\partial
_{0}E_{l(c)}-(E_{(c)}^{~~~l}E_{(a)}^{~~~m}-E_{(a)}^{~~~l}E_{(c)}^{~~~m})D_{l}N_{m}]-%
\mathcal{D}_{i}\omega _{0(0a)}%
\end{array}
\label{40}
\end{equation}%
and%
\begin{equation}
\begin{array}{c}
R_{0i(ab)}=\partial _{0}\omega _{i(ab)}+\left( -E_{(a)}^{~~~j}\partial
_{j}N+N^{j}E_{(a)}^{~~~k}K_{jk}\right) \left( E_{(b)}^{~~~l}K_{il}\right) \\ 
\\ 
-\left( E_{(a)}^{~~~l}K_{il}\right) \left( -E_{(b)}^{~~~j}\partial
_{j}N+N^{j}E_{(b)}^{~~~k}K_{jk}\right) -\mathcal{D}_{i}\omega _{0(ab)}.%
\end{array}
\label{41}
\end{equation}

Let us now consider the scalar curvature tensor

\begin{equation}
R=e_{(A)}^{~~~\mu }e_{(B)}^{~~~\nu }R_{\mu \nu }^{~~~(AB)}.  \label{42}
\end{equation}%
By virtue of (7) we have

\begin{equation}
R=2N^{-1}E_{(a)}^{~~~i}R_{0i}^{~~~(0a)}-2N^{-1}N^{i}E_{(a)}^{~~~j}R_{ij}^{~~~(0a)}+E_{(a)}^{~~~i}E_{(b)}^{~~~j}R_{ij}^{~~~(ab)}
\label{43}
\end{equation}%
or%
\begin{equation}
R=-2N^{-1}E^{(a)i}R_{0i(0a)}+2N^{-1}N^{i}E^{(a)j}R_{ij(0a)}+E_{(a)}^{~~~i}E_{(b)}^{~~~j}R_{ij}^{(ab)}.
\label{44}
\end{equation}%
Therefore, substituting (32), (37), (39) and (40) into (44), we find

\begin{equation}
\begin{array}{c}
R=-N^{-1}\partial _{0}(g_{ij})K^{ij}-2N^{-1}\partial
_{0}(g_{ij}K^{ij})+2N^{-1}E^{(a)i}\mathcal{D}_{i}\omega _{0(0a)} \\ 
\\ 
+2N^{-1}N^{i}E_{(a)}^{~~~j}(\mathcal{D}_{i}(E^{(a)k}K_{jk})-\mathcal{D}%
_{j}(E^{(a)k}K_{ik})) \\ 
\\ 
+E_{(a)}^{~~~i}E_{(b)}^{~~~j}\tilde{R}%
_{ij}^{~~~(ab)}+E_{(a)}^{~~~i}E_{(b)}^{~~~j}\left[
E^{(a)k}E^{(b)l}K_{ik}K_{jl}-E^{(a)k}E^{(b)l}K_{jk}K_{il}\right] ,%
\end{array}
\label{45}
\end{equation}%
where we considered the expression $g^{ij}=E_{(a)}^{~~~i}E^{(a)j}$ and the
property $K_{ij}=K_{ji}$. By using the fact that

\[
\mathcal{D}_{i}E_{j}^{~(a)}=\partial _{i}E_{j}^{~(a)}-\Gamma
_{ij}^{k}(g)E_{k}^{~(a)}+\omega _{i~~(b)}^{~(a)}E_{j}^{~(b)}=0, 
\]%
we find that (45) is reduced to

\begin{equation}
\begin{array}{c}
R=N^{-1}\{-\partial _{0}(g_{ij})K^{ij}-2\partial _{0}(g_{ij}K^{ij})+2%
\mathcal{D}_{i}(E^{i(a)}\omega _{0(0a)}) \\ 
\\ 
+2N^{i}\mathcal{D}_{j}[\delta _{i}^{j}(g^{kl}K_{kl})-g^{jk}K_{ik}]\}+\tilde{R%
}+g^{ij}K_{ij}g^{kl}K_{kl}-K_{ij}K^{ij}.%
\end{array}
\label{46}
\end{equation}%
In this way we see that the action%
\begin{equation}
S_{D}=\int_{M^{D}}\sqrt{-\gamma }R=\int_{M^{D}}\sqrt{g}NR=\int_{M^{D}}\tilde{%
E}NR  \label{47}
\end{equation}%
becomes%
\begin{equation}
\begin{array}{c}
S_{D}=\int_{M^{D}}\tilde{E}\{-\partial _{0}(g_{ij})K^{ij}-2\partial
_{0}(g_{ij}K^{ij}) \\ 
\\ 
-(\mathcal{D}_{j}N_{i}+\mathcal{D}_{j}N_{i})[g^{ij}(g^{kl}K_{kl})-K^{ij}]+N(%
\tilde{R}+g^{ij}K_{ij}g^{kl}K_{kl}-K_{ij}K^{ij}) \\ 
\\ 
+\mathcal{D}_{j}\{+2\tilde{E}\{(E^{j(a)}\omega
_{0(0a)})-N_{i}[g^{ij}(g^{kl}K_{kl})-K^{ij}]\}\},%
\end{array}
\label{48}
\end{equation}%
where $\tilde{E}$ is the determinant of $E_{i}^{~~(a)}$. But according to
(38) we have

\begin{equation}
\mathcal{D}_{j}N_{i}+\mathcal{D}_{i}N_{j}=D_{j}N_{i}+D_{i}N_{j}=2NK_{ij}+%
\partial _{0}(g_{ij}).  \label{49}
\end{equation}%
Thus, up to a surface term (48) yields

\begin{equation}
\begin{array}{c}
S_{D}=\int_{M^{D}}\tilde{E}\{-\partial _{0}(g_{ij})K^{ij}-2\partial
_{0}(g_{ij}K^{ij})-(2NK_{ij} \\ 
\\ 
+\partial _{0}g)[g^{ij}(g^{kl}K_{kl})-K^{ij}]+N(\tilde{R}%
+g^{ij}K_{ij}g^{kl}K_{kl}-K_{ij}K^{ij})\}.%
\end{array}
\label{50}
\end{equation}%
Simplifying this expression we get

\begin{equation}
\begin{array}{c}
S_{D}=\int_{M^{D}}\tilde{E}\{-2\partial _{0}(g_{ij}K^{ij})-\partial
_{0}(g_{ij})g^{ij}(g^{kl}K_{kl}) \\ 
\\ 
+N(\tilde{R}+K_{ij}K^{ij}-g^{ij}K_{ij}g^{kl}K_{kl})\}.%
\end{array}
\label{51}
\end{equation}%
Since $\partial _{0}\tilde{E}=\frac{1}{2}\tilde{E}\partial
_{0}(g_{ij})g^{ij} $ we can further simplify (51) in the form

\begin{equation}
S_{D}=\int_{M^{D}}\{-2\partial _{0}(\tilde{E}g_{ij}K^{ij})+\tilde{E}\{N(%
\tilde{R}+K_{ij}K^{ij}-g^{ij}K_{ij}g^{kl}K_{kl})\}\}.  \label{52}
\end{equation}%
So up to a \ total time derivative we end up with

\begin{equation}
\begin{array}{c}
S_{D}=\int_{M^{D}}L=\int_{M^{D}}\tilde{E}N(\tilde{R}%
+K_{ij}K^{ij}-g^{ij}K_{ij}g^{kl}K_{kl}) \\ 
\\ 
=\int_{M^{D}}\sqrt{g}N(\tilde{R}+K_{ij}K^{ij}-g^{ij}K_{ij}g^{kl}K_{kl}).%
\end{array}
\label{53}
\end{equation}%
This is of course the typical form of the action in canonical gravity (see
Refs. in [17] and references therein).

Let us now introduce the canonical momentum conjugate to $g_{ij}$,

\begin{equation}
\pi ^{ij}=\frac{\partial L}{\partial \partial _{0}g_{ij}}.  \label{54}
\end{equation}%
Using (38) and (53) we obtain

\begin{equation}
\pi ^{ij}=-\tilde{E}(K^{ij}-g^{ij}g^{kl}K_{kl}).  \label{55}
\end{equation}%
Thus, by writing (53) in the form

\begin{equation}
\begin{array}{c}
S_{D}=\int_{M^{D}}\{2\tilde{E}N(K_{ij}K^{ij}-g^{ij}K_{ij}g^{kl}K_{kl}) \\ 
\\ 
+\tilde{E}N\{\tilde{R}-(K_{ij}K^{ij}-g^{ij}K_{ij}g^{kl}K_{kl})\}\}.%
\end{array}
\label{56}
\end{equation}
we see that, in virtue of (55), the first term in (56) can be written as

\begin{equation}
\begin{array}{c}
2\tilde{E}N(K_{ij}K^{ij}-g^{ij}K_{ij}g^{kl}K_{kl})=-2NK_{ij}\pi ^{ij} \\ 
\\ 
=-(-\partial _{0}g_{ij}+D_{i}N_{j}+D_{j}N_{i})\pi ^{ij},%
\end{array}
\label{57}
\end{equation}%
where once again we used (38). Thus, by considering (55) and (57) we find
that up to surface term $S_{D}$ becomes

\begin{equation}
\begin{array}{c}
S_{D}=\int_{M^{D}}\{\partial _{0}g_{ij}\pi ^{ij}+2N_{i}D_{j}\pi ^{ij} \\ 
\\ 
+\tilde{E}N\{\tilde{R}-\frac{1}{\tilde{E}^{2}}(\pi _{ij}\pi ^{ij}-\frac{1}{%
D-2}g^{ij}\pi _{ij}g^{kl}\pi _{kl})\}\}.%
\end{array}
\label{58}
\end{equation}%
We see that $N$ and $N^{i}$ play the role of Lagrange multiplier and
therefore from (58) it follows that the Diffeomorphism and Hamiltonian
constraints are

\begin{equation}
H^{i}\equiv 2D_{j}\pi ^{ij}  \label{59}
\end{equation}%
and

\begin{equation}
H\equiv \tilde{E}\{\tilde{R}-\frac{1}{\tilde{E}^{2}}(\pi _{ij}\pi ^{ij}-%
\frac{1}{D-2}g^{ij}\pi _{ij}g^{kl}\pi _{kl}),  \label{60}
\end{equation}%
respectively. The expression (60) can also be written as

\begin{equation}
H=\sqrt{g}\tilde{R}-\frac{1}{\sqrt{g}}(\pi _{ij}\pi ^{ij}-\frac{1}{D-2}%
g^{ij}\pi _{ij}g^{kl}\pi _{kl}).  \label{61}
\end{equation}%
Even with a rough inspection of the constraint (61) one can expect that "the
potential term" $\tilde{R}$ presents serious difficulties when we make the
transition to the quantum scenario;

\begin{equation}
\hat{H}^{i}\mid \psi >=0  \label{62}
\end{equation}%
and%
\begin{equation}
\hat{H}\mid \psi >=0.  \label{63}
\end{equation}%
We would like to remark that according to our development this is true no
just in four dimensions but in an arbitrary dimension $D$.

\bigskip\ 

\noindent \textbf{3.- Palatini formalism}

\smallskip\ 

Similar conclusion, in relation to the quantization of "the potential term" $%
\tilde{R},$ can be obtained if we use the so called Palatini formalism. In
this case the variables $E_{(A)}^{~~~\mu }$ and $\omega _{\nu }^{~(AB)}$ are
considered as independent variables. We start again with the action (47),
namely $S_{D}=\int_{M^{D}}\tilde{E}NR,$ with $R$ given by (44). Substituting
(32), (34) and (35) into (47) we find%
\begin{equation}
\begin{array}{c}
S_{D}=\int_{M^{D}}\tilde{E}\{-2E^{(a)i}[\partial _{0}\omega
_{i(0a)}-\partial _{i}\omega _{0(0a)}+\omega _{0(0c)}\omega
_{i~~(a)}^{~(c)}-\omega _{i(0c)}\omega _{0~~(a)}^{~(c)}] \\ 
\\ 
+2N^{i}E^{(a)j}[\mathcal{D}_{i}\omega _{j(0a)}-\mathcal{D}_{j}\omega
_{i(0a)}] \\ 
\\ 
+NE^{(a)i}E^{(b)j}[\tilde{R}_{ij(ab)}+\omega _{i(0a)}\omega _{j(0b)}-\omega
_{j(0a)}\omega _{i(0b)}],%
\end{array}
\label{64}
\end{equation}%
which can also be written as

\begin{equation}
\begin{array}{c}
S_{D}=\int_{M^{D}}\{-2\tilde{E}E^{(a)i}\partial _{0}\omega
_{i(0a)}+NE^{(a)i}E^{(b)j}[\tilde{R}_{ij(ab)}+\omega _{i(0a)}\omega _{j(0b)}
\\ 
\\ 
-\omega _{j(0a)}\omega _{i(0b)}]-2\tilde{E}E^{(a)i}\mathcal{D}_{i}\omega
_{0(0a)}+2N^{i}E^{(a)j}[\mathcal{D}_{i}\omega _{j(0a)}-\mathcal{D}_{j}\omega
_{i(0a)}]\}.%
\end{array}
\label{65}
\end{equation}%
The last two terms in (65) can be used for obtaining the formula $\mathcal{D}%
_{i}E_{j}^{~(a)}=0$ as a field equation. So if we focus in the first two
terms in (65) we see that the quantities $\tilde{E}E^{i(a)}$ and $\omega
_{i(0a)}$ can be considered as conjugate canonical variables, with $\tilde{E}%
E^{i(a)}$ playing the role of a conjugate momentum to $\omega _{i(0a)}$,
while the expression%
\begin{equation}
H=E^{(a)i}E^{(a)j}[\tilde{R}_{ij(ab)}+\omega _{i(0a)}\omega _{j(0b)}-\omega
_{j(0a)}\omega _{i(0b)}]  \label{66}
\end{equation}%
plays the role of a Hamiltonian constraint. So when we proceed to quantize
the system we again expect to find some difficulties because of the term $%
\tilde{R}=E_{(a)}^{~~~i}E_{(b)}^{~~~j}\tilde{R}_{ij(ab)}.$ Once again, this
is true in any dimension $D$.

\bigskip\ 

\noindent \textbf{4.- Self-dual formalism in four dimensions}

\smallskip\ 

In four dimensions something interesting happens if instead of (47) one
considers the alternative action [2]-[3]

\begin{equation}
^{+}S_{4}=\frac{1}{2}\int_{M^{4}}ee_{(A)}^{~~~\mu }e_{(B)}^{~~~\nu
}~^{+}R_{\mu \nu }^{~~(AB)}.  \label{67}
\end{equation}%
Here,

\begin{equation}
^{\pm }R_{\mu \nu }^{~~(AB)}=\frac{1}{2}~^{\pm }M_{~~~~~(CD)}^{(AB)}R_{\mu
\nu }^{~~(CD)},  \label{68}
\end{equation}%
with

\begin{equation}
^{\pm }M_{~~~~~(CD)}^{(AB)}=\frac{1}{2}(\delta _{~~~~~(CD)}^{(AB)}\mp
i\epsilon _{~~~~~(CD)}^{(AB)})  \label{69}
\end{equation}%
is the self(anti-self)-dual sector of $R_{\mu \nu }^{(AB)}.$ The symbol $%
\delta _{~~~~~(CD)}^{(AB)}=\delta _{(C)}^{(A)}\delta _{(D)}^{(B)}-\delta
_{(C)}^{(B)}\delta _{(D)}^{(A)}$ denotes a generalized delta. (Observe that
the presence of the completely antisymmetric symbol $\epsilon _{(CD)}^{(AB)}$
in (60) is an indication that the spacetime dimension is equal to four.)
Since $^{+}R_{\mu \nu }^{(AB)}$ is self-dual, that is

\begin{equation}
\frac{1}{2}\epsilon _{~~~~~(CD)}^{(AB)}~^{+}R_{\mu \nu
}^{~~(CD)}=i~^{+}R_{\mu \nu }^{~~(AB)},  \label{70}
\end{equation}%
we find that $^{+}S$ can be written as%
\begin{equation}
\begin{array}{c}
^{+}S_{4}=\frac{1}{2}\int_{M^{4}}E%
\{2E_{(0)}^{~~~0}E_{(a)}^{~~~i}~^{+}R_{0i}^{~~(0a)}+2E_{(0)}^{~~~i}E_{(a)}^{~~~j}~^{+}R_{ij}^{~~(0a)}
\\ 
\\ 
-i\frac{1}{2}E_{(a)}^{~~~i}E_{(b)}^{~~~j}\varepsilon ^{abc}~^{+}R_{ij(0c)}\},%
\end{array}
\label{71}
\end{equation}%
showing that only ${}^{+}R_{\mu \nu }^{~~~(0a)}$ is needed. Here we used the
definition $\epsilon ^{abc}\equiv \epsilon ^{0abc}$. A fine point is that up
to the Bianchi identities for $R_{\mu \nu }^{~~~(AB)},$ $^{+}S_{4}$ is
equivalent to $S_{4}$. If we use the $3+1$ decomposition (6) and (7) we find
that (71) becomes%
\begin{equation}
\begin{array}{c}
^{+}S_{4}=-\int_{M^{4}}\tilde{E}\{2E_{(a)}^{~~~i}~
^{+}R_{0i}^{~~(0a)}-2N^{i}E_{(a)}^{~~~j}~^{+}R_{ij}^{~~(0a)} \\ 
\\ 
-i\frac{1}{2}NE_{(a)}^{~~~i}E_{(b)}^{~~~j}\varepsilon
_{c}^{ab}~^{+}R_{ij}^{~~(0c)}\}.%
\end{array}
\label{72}
\end{equation}%
According to (35), we discover that the first term in (72) establishes that $%
\tilde{E}E_{(a)}^{~~~i}$ can be understood as the canonical momentum
conjugate to $^{+}\omega _{i}^{(0a)}$. Thus one can interpret the second and
the third terms in (64) as the canonical constraints,

\begin{equation}
^{+}H^{i}=-2\tilde{E}E_{(a)}^{~~~j}~^{+}R_{ij}^{~~(0a)}=0  \label{73}
\end{equation}%
and

\begin{equation}
^{+}H=-i\frac{1}{2}\tilde{E}E_{(a)}^{~~~i}E_{(b)}^{~~~j}\varepsilon
^{abc}~^{+}R_{ij(0c)}=0,  \label{74}
\end{equation}%
(see Ref. [42]). Comparing (66) and (74) one sees that the term $\tilde{R}%
=E^{(a)i}E^{(b)j}\tilde{R}_{ij(ab)}$ is not manifest in (74). At first sight
one may expect that this reduced result of the Diffeomorphism and
Hamiltonian constraints may induce a simplification at the quantum level.
However, it is known that there are serious difficulties for finding the
suitable representation for the corresponding associated states with (73)
and (74). This is true, for instance, when one tries to find suitable
representation of the reality condition associated with the connection.

One of the key ingredients to achieve the simpler constraint (74) is, of
course, the self-duality of $^{+}R_{\mu \nu }^{~~(AB)}$. This mechanism
works in four dimensions because of the lemma; the dual of a two form is
another two form. This is, of course, true because we are using the $%
\epsilon $-symbol to define duality. Thus, in higher dimensions this lemma
is no longer true. However, in eight dimensions there exist another
possibility to define duality as we shall see in section 6.

\bigskip\ 

\noindent \textbf{5.- Generalization of self-dual formalism in four
dimensions}

\smallskip\ 

In this section we shall apply the canonical formalism to the action
[18]-[19]

\begin{equation}
\mathcal{S}_{4}=-\frac{1}{16}\int_{M^{4}}\varepsilon ^{\mu \nu \alpha \beta
}~^{+}\mathcal{R}_{\mu \nu }^{~~(AB)}~^{+}\mathcal{R}_{\alpha \beta
}^{~~(CD)}\epsilon _{(ABCD)},  \label{75}
\end{equation}%
which is a generalization of (67). Here,

\begin{equation}
\mathcal{R}_{\mu \nu }^{~~(AB)}=R_{\mu \nu }^{~~(AB)}+\Sigma _{\mu \nu
,}^{~~(AB)}  \label{76}
\end{equation}%
with\ $R_{\mu \nu (AB)}$ defined in (31) and

\begin{equation}
\Sigma _{\mu \nu }^{~~(AB)}=e_{\mu }^{~(A)}e_{\nu }^{~(B)}-e_{\mu
}^{~(B)}e_{\nu }^{~(A)}.  \label{77}
\end{equation}%
In fact, by substituting (76) and (77) into (75) one can show that the
action (75) is reduced to three terms: topological invariant term,
cosmological constant term and the action (67).

By using (70) it is not difficult to see that (75) can be decomposed as

\begin{equation}
\mathcal{S}_{4}=-\frac{i}{2}\int_{M^{4}}\varepsilon ^{\mu \nu \alpha \beta
}~^{+}\mathcal{R}_{\mu \nu }^{~~(0a)}~^{+}\mathcal{R}_{\alpha \beta (0a)}.
\label{78}
\end{equation}%
Further decomposition gives

\begin{equation}
\mathcal{S}_{4}=-i\int_{M^{4}}\varepsilon ^{ijk}~^{+}\mathcal{R}%
_{0i}^{~~(0a)}~^{+}\mathcal{R}_{jk(0a)}.  \label{79}
\end{equation}%
Considering (76) we obtain

\begin{equation}
\begin{array}{c}
\mathcal{S}_{4}=-i\int_{M^{4}}\{\varepsilon
^{ijk}~^{+}R_{0i}^{~~(0a)}~^{+}R_{jk(0a)}+\varepsilon ^{ijk}~^{+}\Sigma
_{0i}^{~~(0a)}~^{+}R_{jk(0a)} \\ 
\\ 
+\varepsilon ^{ijk}~^{+}R_{0i}^{~~(0a)}~^{+}\Sigma _{jk(0a)}+\varepsilon
^{ijk}~^{+}\Sigma _{0i}^{~~(0a)}~^{+}\Sigma _{jk(0a)}\}.%
\end{array}
\label{80}
\end{equation}%
Using (32) and (35) one sees that the first term is a surface term as
expected, while the last term is a cosmological constant term. Thus, by
focusing only in the second and third terms we get

\begin{equation}
^{+}\mathcal{S}_{4}=-i\int_{M^{4}}\{\varepsilon ^{ijk}~^{+}\Sigma
_{0i}^{~~(0a)}~^{+}R_{jk(0a)}+\varepsilon
^{ijk}~^{+}R_{0i}^{~~(0a)}~^{+}\Sigma _{jk(0a)}\},  \label{81}
\end{equation}%
which can be reduced to

\begin{equation}
\begin{array}{c}
^{+}\mathcal{S}_{4}=-i\int_{M^{4}}\{\frac{1}{2}N\varepsilon
^{ijk}~E_{i}^{~(a)}~^{+}R_{jk(0a)}+\frac{i}{2}N^{l}\varepsilon
^{ijk}~\varepsilon _{~~~(bc)}^{(a)}E_{i}^{~(b)}E_{l}^{~(c)}~^{+}R_{jk(0a)}
\\ 
\\ 
-\frac{i}{2}\varepsilon ^{ijk}\varepsilon
_{~~~(bc)}^{(a)}E_{j}^{~(b)}E_{k}^{~(c)}~^{+}R_{0i(0a)}\}.%
\end{array}
\label{82}
\end{equation}%
In turn, it is straightforward to prove that this action reduces to the
action (72). So, the constraints (73) and (74) can also be written as

\begin{equation}
H=-\frac{i}{2}\varepsilon ^{ijk}~E_{i}^{~(a)}~^{+}R_{jk(0a)}=0  \label{83}
\end{equation}%
and

\begin{equation}
H_{l}=\frac{1}{2}\varepsilon ^{ijk}~\varepsilon
_{~~~(bc)}^{(a)}E_{i}^{~(b)}E_{l}^{~(c)}~^{+}R_{jk(0a)}=0.  \label{84}
\end{equation}%
It is interesting to observe the simplicity of the present construction in
contrast to the development of sections 3 and 4.

\bigskip\ 

\noindent \textbf{6. Self-dual formalism in eight dimensions}

\smallskip\ 

One of the key ingredients for achieving the simpler route in the derivation
of the constraints (83) and (84) is, of course, the self-duality of $%
^{+}R_{\mu \nu }^{~~(AB)}$. This works in four dimensions because the dual
of a two form is another two form. However, in higher dimensions this line
of though is difficult to sustain except in eight dimensions. In fact, one
can attempt to generalize the formalism of section 4 to higher dimensions
using BF technics [22] but the self-dual property is lost as it was
described in section 4. On the other hand in eight dimensions one may take
recourse of the octonionic structure constants and define a self-dual four
form $\eta ^{\mu \nu \alpha \beta }$ which can be used to construct similar
approach to the one presented in section 4 as it was proved in Refs. [13]
and [14]. The aim of this section is to pursuing this idea by exploring the
possibility of bringing the formalism to the quantum scenario.

Our starting point is the action [13]

\begin{equation}
\mathcal{S}_{8}=\frac{1}{192}\int_{M^{8}}e\eta ^{\mu \nu \alpha \beta }~^{+}%
\mathcal{R}_{\mu \nu }^{~~(AB)}~^{+}\mathcal{R}_{\alpha \beta }^{~~(CD)}\eta
_{(ABCD)}.  \label{85}
\end{equation}%
Here, the indices $\mu ,\nu ,..etc$ are "spacetime" indices, running from $0$
to $7$, while the indices $A,B,..etc$ are frame indices running also from $0$
to $7$. (Just by convenience in what follows, we shall assume an Euclidean
signature.) The quantity $e$ is the determinant of the eight dimensional
matrix $e_{\mu }^{~(A)}$.

In addition, we have the following definition:

\begin{equation}
\mathcal{R}_{\mu \nu }^{~~(AB)}=R_{\mu \nu }^{~~(AB)}+\Sigma _{\mu \nu
,}^{~~(AB)}  \label{86}
\end{equation}%
with\ 

\begin{equation}
R_{\mu \nu (AB)}=\partial _{\mu }\omega _{\nu (AB)}-\partial _{\nu }\omega
_{\mu (AB)}+\omega _{\mu (AC)}\omega _{\nu ~~(B)}^{~(C)}-\omega _{\mu
(BC)}\omega _{\nu ~~(A)}^{~(C)}  \label{87}
\end{equation}%
and

\begin{equation}
\Sigma _{\mu \nu }^{~~(AB)}=e_{\mu }^{~(A)}e_{\nu }^{~(B)}-e_{\mu
}^{~(B)}e_{\nu }^{~(A)}.  \label{88}
\end{equation}%
The $\eta $-symbol $\eta _{(ABCD)}$ is a completely antisymmetric object,
which is related with the octonion structure constants $\eta _{(abc0)}=\psi
_{abc}$ and its dual $\eta _{(abcd)}=\varphi _{(abcd)}$, satisfying the
self-dual (anti-self-dual) formula

\begin{equation}
\eta _{(ABCD)}=\frac{\varsigma }{4!}\varepsilon _{(ABCDEFGH)}\eta ^{(EFGH)}.
\label{89}
\end{equation}%
For $\varsigma =1,$ $\eta _{(ABCD)}$ is self-dual (and for $\varsigma =-1$
is anti-self-dual). Moreover, $\eta $-symbol satisfies the relations
[20]-[21] (see also Refs. [5] and [6]),

\begin{equation}
\eta _{(ABCD)}\eta ^{(EFCD)}=6\delta _{~~~~~(AB)}^{(EF)}+4\eta
_{~~~~~(AB)}^{(EF)},  \label{90}
\end{equation}%
\begin{equation}
\eta _{(ABCD)}\eta ^{(EBCD)}=42\delta _{A}^{E},  \label{91}
\end{equation}%
and%
\begin{equation}
\eta _{(ABCD)}\eta ^{(ABCD)}=336.  \label{92}
\end{equation}%
Finally, by introducing the dual of $\mathcal{R}_{\mu \nu }^{~~(AB)}$ in the
form

\begin{equation}
^{\star }\mathcal{R}_{\mu \nu }^{(AB)}=\frac{1}{2}\eta _{~~~~~(CD)}^{(AB)}%
\mathcal{R}_{\mu \nu }^{~~(CD)},  \label{93}
\end{equation}%
we define the self-dual $^{+}\mathcal{R}_{\mu \nu }^{~~(AB)}$ and
anti-self-dual $^{-}\mathcal{R}_{\mu \nu }^{~~(AB)}$ parts of $\mathcal{R}%
_{\mu \nu }^{~~(AB)}$ in the form

\begin{equation}
^{+}\mathcal{R}_{\mu \nu }^{~~(AB)}=\frac{1}{4}(\mathcal{R}_{\mu \nu
}^{~~(AB)}+^{\star }\mathcal{R}_{\mu \nu }^{~~(AB)})  \label{94}
\end{equation}%
and%
\begin{equation}
^{-}\mathcal{R}_{\mu \nu }^{~~(AB)}=\frac{1}{4}(3\mathcal{R}_{\mu \nu
}^{~~(AB)}-^{\star }\mathcal{R}_{\mu \nu }^{~~(AB)}),  \label{95}
\end{equation}%
respectively. Since

\begin{equation}
^{\star \star }\mathcal{R}_{\mu \nu }^{~~(AB)}=3\mathcal{R}_{\mu \nu
}^{~~(AB)}+2^{\star }\mathcal{R}_{\mu \nu }^{~~(AB)},  \label{96}
\end{equation}%
we see that

\begin{equation}
^{\star +}\mathcal{R}_{\mu \nu }^{~~(AB)}=3^{+}\mathcal{R}_{\mu \nu
}^{~~(AB)}  \label{97}
\end{equation}%
and

\begin{equation}
^{\star -}\mathcal{R}_{\mu \nu }^{~~(AB)}=-^{-}\mathcal{R}_{\mu \nu
}^{~~(AB)}.  \label{98}
\end{equation}%
Thus, up to a numerical factor we see that $^{+}\mathcal{R}_{\mu \nu
}^{~~(AB)}$ and $^{-}\mathcal{R}_{\mu \nu }^{~~(AB)}$ play, in fact, the
role of the self-dual and anti-self-dual parts, respectively of $\mathcal{R}%
_{\mu \nu }^{~~(AB)}.$ It turns out to be convenient to write (94) as [12]

\begin{equation}
^{+}\mathcal{R}_{\mu \nu }^{~~(AB)}=\frac{1}{2}~^{+}\Lambda
_{~~~~~(CD)}^{(AB)}\mathcal{R}_{\mu \nu }^{~~(CD)},  \label{99}
\end{equation}%
where

\begin{equation}
^{+}\Lambda _{~~~~~(CD)}^{(AB)}=\frac{1}{4}(\delta _{~~~~~(CD)}^{(AB)}+\eta
_{~~~~~(CD)}^{(AB)}).  \label{100}
\end{equation}%
While, (95) can be written in the form

\begin{equation}
^{-}\mathcal{R}_{\mu \nu }^{~~(AB)}=\frac{1}{2}~^{-}\Lambda
_{~~~~~(CD)}^{(AB)}\mathcal{R}_{\mu \nu }^{~~(CD)},  \label{101}
\end{equation}%
with

\begin{equation}
^{-}\Lambda _{~~~~~(CD)}^{(AB)}=\frac{1}{4}(3\delta _{~~~~~(CD)}^{(AB)}-\eta
_{~~~~~(CD)}^{(AB)}).  \label{102}
\end{equation}%
The objects $^{\pm }\Lambda $ admit an interpretation of projection
operators. In fact, one can prove that the objects $^{+}\Lambda $ and $%
^{-}\Lambda ,$ given in (100) and (102) respectively, satisfy [12]

\begin{equation}
^{+}\Lambda +^{-}\Lambda =1,  \label{103}
\end{equation}

\begin{equation}
^{+}\Lambda ^{-}\Lambda =^{-}\Lambda ^{+}\Lambda =0,  \label{104}
\end{equation}

\begin{equation}
^{+}\Lambda ^{2}=^{+}\Lambda ,  \label{105}
\end{equation}%
and

\begin{equation}
^{-}\Lambda ^{2}=^{-}\Lambda .  \label{106}
\end{equation}%
Here, $^{\pm }\Lambda ^{2}$ means $\frac{1}{4}^{\pm }\Lambda
_{~~~~~(CD)}^{(AB)\pm }\Lambda _{~~~~~(GH)}^{(EF)}\delta _{(ABEF)}$.

Finally, the object $\eta ^{\mu \nu \alpha \beta }$ is a completely
antisymmetric tensor determined by the relation

\begin{equation}
\eta _{\mu \nu \alpha \beta }\equiv e_{\mu }^{(A)}e_{\nu }^{(B)}e_{\alpha
}^{(C)}e_{\beta }^{(D)}\eta _{(ABCD)}.  \label{107}
\end{equation}

Before we explore the consequences of (85) let us try to understand the
volume element structure in (85) from alternative analysis. For this purpose
it turns out convenient to define the quantity

\begin{equation}
\hat{e}\equiv \frac{1}{4!}\hat{\eta}^{\mu \nu \alpha \beta }e_{\mu
}^{(A)}e_{\nu }^{(B)}e_{\alpha }^{(C)}e_{\beta }^{(D)}\eta _{(ABCD)},
\label{108}
\end{equation}%
where, $\hat{\eta}^{\mu \nu \alpha \beta }$ takes values in the set $%
\{-1,0,1\}$ and has exactly the same octonionic properties as $\eta
_{(ABCD)} $ (specified in (89)-(92)). The formula (108) can be understood as
the analogue of the determinant for $e_{\mu }^{(A)}$ in four dimensions.
Thus, by using the octonionic properties (89)-(92) for $\eta _{(ABCD)},$
such as the self-duality relation

\begin{equation}
\eta ^{(ABCD)}=\frac{1}{4!}\varepsilon ^{(ABCDEFGH)}\eta _{(EFGH)},
\label{109}
\end{equation}%
from (107) one can prove that up to numerical constants $a=\frac{1}{5}$ and $%
b=\frac{1}{3}$ one obtains

\begin{equation}
\hat{e}\eta ^{\mu \nu \alpha \beta }=a\hat{\eta}^{\mu \nu \alpha \beta }+b%
\hat{\eta}^{\mu \nu \tau \lambda }\eta _{\tau \lambda }^{\alpha \beta },
\label{110}
\end{equation}%
which proves that at least $\hat{\eta}^{\mu \nu \alpha \beta }\sim \hat{e}%
\eta ^{\mu \nu \alpha \beta }.$ The expression (110) means that there are
two terms in (85), one which can be written as

\begin{equation}
\mathcal{S}_{8}\sim \frac{1}{192}\int_{M^{8}}\frac{e}{\hat{e}}\hat{\eta}%
^{\mu \nu \alpha \beta }~^{+}\mathcal{R}_{\mu \nu }^{~~(AB)}~^{+}\mathcal{R}%
_{\alpha \beta }^{~~(CD)}\eta _{(ABCD)}.  \label{111}
\end{equation}%
In four dimensions the corresponding ratio $\frac{e}{\hat{e}}$ gives $\frac{e%
}{\hat{e}}=1$. However, the situation is more subtle in eight dimensions
because we can not set $\frac{e}{\hat{e}}=1$ and this suggests an exotic
volume element mediated in part by the exceptional group $G_{2}$. This is
suggested in part because the quantities $\hat{\eta}^{\mu \nu \alpha \beta }$
and $\eta _{(ABCD)}$ are only $G_{2}$-invariant rather than $SO(8)$%
-invariant.

Now considering (107) and (109) one observes that $\eta ^{\mu \nu \alpha
\beta }$ is also self-dual in eight dimensions, that is

\begin{equation}
\eta ^{\mu \nu \alpha \beta }=\frac{1}{4!}\epsilon ^{\mu \nu \alpha \beta
\lambda \rho \sigma \tau }\eta _{\lambda \rho \sigma \tau },  \label{112}
\end{equation}%
which implies that the action (85) can also be written as

\begin{equation}
\mathcal{S}_{8}=\frac{1}{(192)4!}\int d^{8}x~e~\epsilon ^{\lambda \rho
\sigma \tau \mu \nu \alpha \beta }\eta _{\lambda \rho \sigma \tau }~^{+}%
\mathcal{R}_{\mu \nu }^{~~(AB)}~^{+}\mathcal{R}_{\alpha \beta }^{~~(CD)}\eta
_{(ABCD)}  \label{113}
\end{equation}%
or

\begin{equation}
\mathcal{S}_{8}=\frac{1}{(192)4!}\int d^{8}x~\varepsilon ^{\mu \nu \alpha
\beta \lambda \rho \sigma \tau }\eta _{\lambda \rho \sigma \tau }~^{+}%
\mathcal{R}_{\mu \nu }^{~~(AB)}~^{+}\mathcal{R}_{\alpha \beta }^{~~(CD)}\eta
_{(ABCD)},  \label{114}
\end{equation}%
since%
\begin{equation}
\epsilon ^{\mu \nu \alpha \beta \lambda \rho \sigma \tau }=\frac{1}{e}%
\varepsilon ^{\mu \nu \alpha \beta \lambda \rho \sigma \tau }.  \label{115}
\end{equation}%
Here, we recall that the quantity $e$ denotes the usual determinant of $%
e_{\mu }^{(A)}$ in eight dimensions. The expression (114) allows us to write
(85) in the alternative form

\begin{equation}
\mathcal{S}_{8}=\frac{1}{(192)4!}\int_{M^{8}}~\eta \wedge ~^{+}\mathcal{R}%
^{~~(AB)}\wedge ~^{+}\mathcal{R}^{~~(CD)}\eta _{(ABCD)}.  \label{116}
\end{equation}

Now, since

\begin{equation}
^{+}\mathcal{R}_{\mu \nu }^{~~(AB)}=~^{+}R_{\mu \nu }^{~~(AB)}+~^{+}\Sigma
_{\mu \nu }^{~~(AB)},  \label{117}
\end{equation}%
one finds that the action (85) becomes

\begin{equation}
\mathcal{S}_{8}=\frac{1}{192}\int_{M^{8}}e(T+K+C),  \label{118}
\end{equation}%
with

\begin{equation}
T=\eta ^{\mu \nu \alpha \beta }~^{+}R_{\mu \nu }^{~~(AB)}~^{+}R_{\alpha
\beta }^{~~(CD)}\eta _{(ABCD)},  \label{119}
\end{equation}

\begin{equation}
K=2\eta ^{\mu \nu \alpha \beta }~^{+}\Sigma _{\mu \nu
}^{~~(AB)}~^{+}R_{\alpha \beta }^{~~(CD)}\eta _{(ABCD)},  \label{120}
\end{equation}%
and

\begin{equation}
C=\eta ^{\mu \nu \alpha \beta }~^{+}\Sigma _{\mu \nu }^{~~(AB)}~^{+}\Sigma
_{\alpha \beta }^{~~(CD)}\eta _{(ABCD)}.  \label{121}
\end{equation}%
It turns out that the $T$ term can be identified with a topological
invariant in eight dimensions. In fact, it can be considered as the
"gravitational" analogue of the topological term of $G_{2}$-invariant super
Yang-Mills theory [23];

\begin{equation}
\mathcal{S}_{YM}=\int_{M^{8}}\eta ^{\mu \nu \alpha \beta }F_{\mu \nu
}^{~~a}F_{\alpha \beta }^{~~b}g_{ab},  \label{122}
\end{equation}%
where $F_{\mu \nu }^{a}$ is the Yang-Mills field strength and $g_{ab}$ is
the group invariant metric. Similarly, $K$ should lead to a kind of gravity
in eight dimensions. Finally, $C$ may be identified with the analogue of a
cosmological constant term. It is worth mentioning that, in general, the $%
\epsilon $-symbol is Lorentz invariant in any dimension, but in contrast the 
$\eta $-symbol is only $SO(7)$-invariant and therefore one must have that
the action (85) is only $SO(7)$-invariant.

For our purpose we shall focus in the $K$-sector of (118), namely

\begin{equation}
^{+}\mathcal{S}_{8}=\frac{1}{96}\int_{M^{8}}e\eta ^{\mu \nu \alpha \beta
}~^{+}\Sigma _{\mu \nu }^{~~(AB)}~^{+}R_{\alpha \beta }^{~~(CD)}\eta
_{(ABCD)},  \label{123}
\end{equation}%
which in virtue of (97) can also be written as

\begin{equation}
^{+}\mathcal{S}_{8}=\frac{1}{16}\int_{M^{8}}e\eta ^{\mu \nu \alpha \beta
}~^{+}\Sigma _{\mu \nu }^{~~(AB)}~^{+}R_{\alpha \beta (AB)}.  \label{124}
\end{equation}

We are ready to develop a canonical decomposition of (124). We get

\begin{equation}
^{+}\mathcal{S}_{8}=\frac{1}{16}\int_{M^{8}}e\{2\eta ^{\mu \nu \alpha \beta
}~^{+}\Sigma _{\mu \nu }^{~~(0a)}~^{+}R_{\alpha \beta (0a)}+\eta ^{\mu \nu
\alpha \beta }~^{+}\Sigma _{\mu \nu }^{~~(ab)}~R_{\alpha \beta (ab)}\},
\label{125}
\end{equation}%
which can be written as%
\begin{equation}
^{+}\mathcal{S}_{8}=\frac{1}{2}\int_{M^{8}}e\eta ^{\mu \nu \alpha \beta
}~^{+}\Sigma _{\mu \nu }^{~~(0a)}~^{+}R_{\alpha \beta (0a)}.  \label{126}
\end{equation}%
Here we used the property $\eta _{(0acd)}\eta ^{(0bcd)}=\eta _{(acd)}\eta
^{(bcd)}=\psi _{acd}\psi ^{bcd}=6\delta _{a}^{b}$, which can be derived from
(80), and we considered the fact that $~~^{+}R_{\alpha \beta (bc)}=\eta
_{~~~(bc)}^{(a)}\ ^{+}R_{\alpha \beta (0a)}$. A further decomposition of
(126) gives

\begin{equation}
^{+}\mathcal{S}_{8}=\int_{M^{8}}\tilde{E}\{\eta ^{ijk}~^{+}\Sigma
_{0i}^{~~(0a)}~^{+}R_{jk(0a)}+\eta ^{ijk}~^{+}\Sigma
_{ij}^{~~(0a)}~^{+}R_{0k(0a)}\},  \label{127}
\end{equation}%
which can be reduce to

\begin{equation}
\begin{array}{c}
^{+}\mathcal{S}_{8}=\int_{M^{8}}\tilde{E}\{\frac{1}{4}N\eta
^{ijk}~E_{i}^{~(a)}~^{+}R_{jk(0a)}+\frac{1}{4}N^{l}\eta ^{ijk}~\eta
_{~~~(bc)}^{(a)}E_{i}^{~(b)}E_{l}^{~(c)}~^{+}R_{jk(0a)} \\ 
\\ 
+\frac{1}{4}\eta ^{ijk}\eta
_{(bca)}E_{j}^{~(b)}E_{k}^{~(c)}~^{+}R_{0i}^{(0a)}\}.%
\end{array}
\label{128}
\end{equation}%
So, the constraints derived from the action (128) are

\begin{equation}
\mathcal{H}=\frac{1}{4}\tilde{E}\eta ^{ijk}~E_{i}^{~(a)}~^{+}R_{jk(0a)}=0
\label{129}
\end{equation}%
and

\begin{equation}
\mathcal{H}_{l}=\frac{1}{4}\tilde{E}\eta ^{ijk}~\eta
_{~~~(bc)}^{(a)}E_{i}^{~(b)}E_{l}^{~(c)}~^{+}R_{jk(0a)}=0.  \label{130}
\end{equation}%
Observe that the term $\tilde{R}=E^{(a)i}E^{(b)j}\tilde{R}_{ij(ab)}$ is not
manifest in (129) and therefore, once again, one may expect some
simplification at the quantum level. Therefore, this shows that the
introduction of the self-dual curvature tensor $^{+}R_{\mu \nu }^{~~(AB)}$
using the $\eta $-symbol makes sense in eight dimensions. However, once
again, this possible quantum simplification is an illusion because the need
of the reality condition for the connection may lead to some difficulties
for finding suitable representation which implements such a reality
condition.

One may wonder whether the same construction may be achieved by considering
the anti-self-dual sector via the anti-self-dual curvature tensor $%
^{-}R_{\mu \nu }^{~~(AB)}$. In order to give a possible answer to this
question one requires to analyze the formalism from the perspective of
octonionic representations of the group $SO(8)$. Let us first recall the
case of four dimensions in connection with the norm group of the
quaternions, namely $SO(4)$. In this case one has the decomposition 
\begin{equation}
SO(4)=S^{3}\times S^{3},  \label{131}
\end{equation}%
which, in turn, allows the result

\begin{equation}
\lbrack ^{+}J_{(AB)},^{-}J_{(AB)}]=0,  \label{132}
\end{equation}%
where $^{\pm }J_{(AB)}$ are the self-dual and anti-self-dual components of
the generator $J_{(AB)}$ of $SO(4).$ As a consequence of this one has the
splitting%
\begin{equation}
R_{\mu \nu }^{~~(AB)}=~^{+}R_{\mu \nu }^{~~(AB)}(^{+}\omega )+~^{-}R_{\mu
\nu }^{~~(AB)}(^{-}\omega ).  \label{133}
\end{equation}%
This means that there is not mixture between the self-dual and
anti-self-dual components of $R_{\mu \nu }^{~~(AB)}$ and consequently one
may choose to work either with the self-dual sector or anti-self-dual sector
of $R_{\mu \nu }^{~~(AB)}$.

The case of eight dimensions is more subtle because the decomposition $^{\pm
}R_{\mu \nu }^{~~(AB)}$of $R_{\mu \nu }^{~~(AB)}$, according to the
expressions (94) and (95), is connected to the splitting of the 28
independents generators $J_{(AB)}$ of $SO(8)$ in 7 generators $%
_{R}^{+}J_{(AB)}\equiv (^{+}\Lambda J)_{(AB)}$ and 21 generators $%
_{R}^{-}J_{(AB)}\equiv (^{-}\Lambda J)_{(AB)}$ which do not commute, that
is, the generators $_{R}^{+}J_{(AB)}$ and $_{R}^{-}J_{(AB)}$, corresponding
to $S_{R}^{7}\equiv $ $SO(8)/SO(7)_{R}$ and $SO(7)_{R}$ respectively do not
satisfy the expression (132). In turn, this means that we can not write $%
R_{\mu \nu }^{~~(AB)}$ as in (133). The situation can be saved by
considering beside the right sector, $S_{R}^{7}$ and $SO(7)_{R}$,
corresponding to the value $\varsigma =1$ in the expression (89), the left
sector $S_{L}^{7}\equiv $ $SO(8)/SO(7)_{L}$ and $SO(7)_{L}$ corresponding to
the value $\varsigma =-1$ in (89). In fact, with this tools at hand one
finds the possibility to combine the generators $_{R}^{+}J_{(AB)}$ and $%
_{L}^{+}J_{(AB)}$ of $S_{R}^{7}$ and $S_{L}^{7}$ respectively, rather than $%
_{R}^{+}J_{(AB)}$ and $_{R}^{-}J_{(AB)}$ or $_{L}^{+}J_{(AB)}$ and $%
_{L}^{-}J_{(AB)}$, according to the $SO(8)$-decomposition

\begin{equation}
SO(8)=S_{R}^{7}\times S_{L}^{7}\times G_{2},  \label{134}
\end{equation}%
which is a closer decomposition to (131) (see [12] for details). In this
case the analogue of (133) will be

\begin{equation}
R_{\mu \nu }^{~~(AB)}=~_{R}^{+}R_{\mu \nu }^{~~(AB)}(_{R}^{+}\omega
)+~_{L}^{+}R_{\mu \nu }^{~~(AB)}(_{L}^{+}\omega ),  \label{135}
\end{equation}%
modulo the exceptional group $G_{2}$. We should mention that just by
convenience in our formalism above we wrote $_{R}^{+}R_{\mu \nu }^{~~(AB)}$
as $^{+}R_{\mu \nu }^{~~(AB)}$, but in general it is necessary to keep in
mind the distinction between $_{R}^{+}R_{\mu \nu }^{~~(AB)}(_{R}^{+}\omega )$
and $_{L}^{+}R_{\mu \nu }^{~~(AB)}(_{L}^{+}\omega )$. What it is important
is that one may choose to work either with the $_{R}^{+}R_{\mu \nu
}^{~~(AB)}(_{R}^{+}\omega )$ sector or $_{L}^{+}R_{\mu \nu
}^{~~(AB)}(_{L}^{+}\omega )$ sector of $R_{\mu \nu }^{~~(AB)}$ in the group
manifold $SO(8)/G_{2}$.

\bigskip\ 

\noindent \textbf{7. Toward a background independent quantum gravity in
eight dimensions and final comments}

\smallskip\ 

Having the canonical constraints (129) and (130) we become closer to our
final goal of developing quantum gravity in eight dimensions. In fact in
this section we shall outline possible quantum physical states $\mid \Psi >$
associated with the corresponding Hamiltonian operators $\mathcal{H}^{\prime
}$ and $\mathcal{H}_{l}^{\prime }$ (associated with (129) and (130)
respectively) via the expressions

\begin{equation}
\mathcal{H}^{\prime }\mid \Psi >=0  \label{136}
\end{equation}%
and

\begin{equation}
\mathcal{H}_{l}^{\prime }\mid \Psi >=0.  \label{137}
\end{equation}%
Of course, even from the beginning one may have the feeling that the
physical solutions of (136) and (137) will be more subtle than in the case
of four dimensions. This is in part due to the fact that the topology in
eight dimensions is less understood that in three or four dimensions.
Nevertheless some progress in this direction has been achieved [24].

In order to describe the physical states, which solves (136) and (137), one
may first write the canonical commutations relations:%
\begin{equation}
\begin{array}{c}
\lbrack \hat{A}_{i}^{~(a)}(x),\hat{A}_{j}^{~(b)}(y)]=0, \\ 
\\ 
\lbrack \hat{E}_{(a)}^{~~~i}(x),\hat{E}_{(b)}^{~~~j}(y)]=0, \\ 
\\ 
\lbrack \hat{E}_{(a)}^{~~~i}(x),\hat{A}_{j}^{~(b)}(y)]=\delta _{j}^{i}\delta
_{a}^{b}\delta ^{7}(x,y).%
\end{array}
\label{138}
\end{equation}%
Here, we have made the symbolic transition $^{+}\omega
_{i}^{~(0a)}\rightarrow A_{i}^{~(a)}$ and consider $A_{i}^{~(a)}$ as a $%
spin(7)$ gauge field. We choose units such that $\hbar =1$. It is worth
mentioning that by introducing the analogue generalized determinant (107)
for $E_{i}^{~(a)}$ one may write the conjugate momentum $\hat{E}%
_{(a)}^{~~~i}(x)$ explicitly in terms of $\hat{E}_{i}^{~(a)}$. The next step
is to choose a representation for the operators $\hat{A}_{i}^{~(a)}$ and $%
\hat{E}_{(a)}^{~~~i}$ of the form

\begin{equation}
\begin{array}{c}
\hat{A}_{i}^{~(a)}\Psi (A)=A_{i}^{~(a)}\Psi (A), \\ 
\\ 
\hat{E}_{(a)}^{~~~i}\Psi (A)=\frac{\delta \Psi (A)}{\delta A_{i}^{~(a)}}.%
\end{array}
\label{139}
\end{equation}%
Using these relations one discover that the quantum constraints can be
solved by Wilson loops wave functions

\begin{equation}
\Psi _{\gamma }(A)=trP\exp \int_{\gamma }A  \label{140}
\end{equation}%
labelled by the loops $\gamma $.

Of course these quantum steps are completely analogue to the case of four
dimensions [25]-[27]. However they are necessary if one wants to go forward
in our quantum program. We believe that interesting aspects in this process
can arise if one look for a physical states in terms of the analogue of the
Chern-Simons states in four dimensions. The reason is because Chern-Simons
theory is linked to instantons in four dimensions via the topological term $%
\int_{M^{4}}tr\varepsilon ^{\mu \nu \alpha \beta }F_{\mu \nu }F_{\alpha
\beta }$, while in eight dimensions the topological term should be of the
form $\int_{M^{4}}tr\eta ^{\mu \nu \alpha \beta }F_{\mu \nu }F_{\alpha \beta
}$. Surprisingly this kind of topological terms have already been considered
in the literature in connection with $G_{2}$-instantons (see [23] and
references therein).

The present work just describes the first steps towards the construction of
background independent quantum gravity in eight dimensions. We certainly may
have in the route many of the problems of the traditional Ashtekar formalism
in four dimensions such as the issue of time. However one of the advantage
that may emerge from the present formalism is the possibility to bring many
new ideas from twelve dimensions via the transition $10+2\rightarrow
(3+1)+(7+1)$ [28]$.$ In fact twelve dimensions is one of the most
interesting proposals for building $M$-theory [29]. An example of this,
Smolin [30]-[31] (see also Refs \ [32] and [33]) has described the
possibility to construct background independent quantum gravity in the
context of topological $M$-theory by obtaining Hitchin's 7 seven dimensional
theory, which in principle seems to admit background independent
formulation, from the classical limit of $M$-theory, namely eleven
dimensional supergravity. The idea is focused on an attempt of reducing the
eleven dimensional manifold $M^{1+10}$ in the form

\begin{equation}
M^{1+10}\rightarrow R\times \Sigma \times S^{1}\times R^{3}.  \label{141}
\end{equation}%
Here, $\Sigma $ is a complex six-dimensional manifold. Considering that the
only degree of freedom is the gauge field three form $A$ which is pure gauge 
$A=d\beta $ and therefore locally trivial $dA=0$, the Smolin's conjecture is
that the Hitchin's action can be derived from the lowest dimensional term
that can be made from $d\beta $ on $R\times \Sigma $ of the corresponding
effective action (see Ref. [30] for details). Observing that $\Sigma \times
S^{1}$ is a seven dimensional manifold and since, via the octonion
structure, the solution $0+8$ is related to the seven sphere solution of
eleven dimensional supergravity one is motivated to conjecture that there
must be a connection between our approach of incorporating Ashtekar
formalism in the context of $M$-theory and the Smolin's program. In turn, $M$%
-theory has motivated the study of many mathematical structures such as
oriented matroid theory [34] (see Refs [35]-[39]). Thus we see as
interesting physical possibility a connection between matroid theory and
Ashtekar formalism. The reason for this is that symbols $\varepsilon ^{\mu
\nu \alpha \beta }$ and $\eta ^{\mu \nu \alpha \beta }$ may be identified
with two examples of four rank chirotopes [40] and therefore it is necessary
to find a criterion for the uniqueness of these symbols from these
perspective [41].

Finally, so far in this article we have focused on the Euclidean case via
the possible representations for $SO(8)$. For further research it may be
interesting to investigate the Lorenzian case associated with the group $%
SO(1,7)$. Since $SO(7)$ is a subgroup of $SO(1,7)$ one finds that (up to
some modified numerical factors) most of the algebraic relations for
octonions given in (89)-(92) are similar. For instance, the self-duality
relation (89) should be modified with $\varsigma =\pm i$ instead of $%
\varsigma =\pm 1$. Thus, the discussion at the end of section 6 should be
slightly modified. However, the transition from Euclidean to Lorenzian
signature at the level of the action (85), and its corresponding quantum
theory, may be more complicated. In this case the usual Wick rotation may be
not enough procedure as in canonical gravity in four dimensions [43] and
therefore it may be necessary to consider a modified action with free
parameters controlling the signature of the spacetime.

\bigskip\ 

\noindent \textbf{Acknowledgments: }I would like to thank A. Ashtekar, M.
Bojowald, P. Laguna, A. Corichi and J. Lewandowski for helpful comments and
the Institute of Gravitational for Physics and Geometry at Penn State
University for the hospitality, where part of this work was developed.

\smallskip\


\begin{thebibliography}{99}
\bibitem{1} A. Ashtekar and J. Lewandowski, Class. Quant. Grav. \textbf{21},
R53 (2004); gr-qc/0404018.

\bibitem{2} T. Jacobson and L. Smolin, Class. Quant. Grav. \textbf{5}, 583
(1988).

\bibitem{3} J. Samuel, Pramana J. Phys. \textbf{28}, L429 (1987).

\bibitem{4} A.Ashtekar, Phys. Rev. Lett. \textbf{57}, 2244 (1986).

\bibitem{5} E. Corrigan, C. Devchand, D.B. Fairlie and J. Nuyts, Nucl. Phys.
B \textbf{214}, 452 (1983).

\bibitem{6} K. S. Abdel-Khalek, "Ring division algebras, self duality and
supersymmetry", Ph.D. Thesis (Advisor: Pietro Rotelli) (2000);
hep-th/0002155.

\bibitem{7} H. Nishino and S. Rajpoot, JHEP \textbf{0404}, 020 (2004);
hep-th/0210132.

\bibitem{8} H. Nishino and S. Rajpoot, Phys. Lett. B \textbf{564}, 269
(2003); hep-th/0302059.

\bibitem{9} J. C. Baez, Bull. Amer. Math. Soc. \textbf{39}, 145 (2002).

\bibitem{10} J. A. Nieto and L. N. Alejo-Armenta, Int. J. Mod. Phys. A 
\textbf{16}, 4207 (2001); hep-th/0005184.

\bibitem{11} A. R. Dundarer, F. Gursey and C. H. Tze, J. Math. Phys. \textbf{%
25}, 1496 (1984).

\bibitem{12} A. R. Dundarer and F. Gursey, J. Math. Phys. \textbf{32}, 1178
(1991).

\bibitem{13} J. A. Nieto, Class. Quant. Grav., \textbf{22}, 947 (2005);
hep-th/0410260.

\bibitem{14} J. A. Nieto, Class. Quant. Grav. \textbf{23}, 4387 (2006);
hep-th/0509169.

\bibitem{15} J. A. Nieto, Gen. Rel. Grav. \textbf{39}, 1109 (2007);
hep-th/0506253.

\bibitem{16} H. Nicolai, H. J. Matschull, J. Geom. Phys.\textbf{11}, 15
(1993).

\bibitem{17} T. Thiemann, "Introduction to modern canonical quantum general
relativity", gr-qc/0110034; S. Mercuri and G. Montani, Int. J. Mod. Phys. D 
\textbf{13}, 165 (2004); gr-qc/0310077.

\bibitem{18} J. A. Nieto, O. Obreg\'{o}n and J. Socorro, Phys. Rev. D 
\textbf{50}, R3583 (1994); gr-qc/9402029.

\bibitem{19} J. A. Nieto, Mod. Phys. Lett. A \textbf{20}, 2157 (2005);
hep-th/0411124.

\bibitem{20} M. Gunaydin and Gursey, J. Math. Phys. \textbf{14}, 1651 (1973).

\bibitem{21} K. Sfetsos, Nucl. Phys. B \textbf{629}, 417 (2002);
hep-th/0112117.

\bibitem{22} L. Freidel, K. Krasnov and R. Puzio, Adv. Theor. Math. Phys. 
\textbf{3}, 1289 (1999); hep-th/9901069.

\bibitem{23} D. M\"{u}lsch and B. Geyer, Int. J. Geom. Meth. Mod. Phys. 
\textbf{1}, 185 (2004); hep-th/0310237.

\bibitem{24} D. Joyce, J. Diff. Geom. \textbf{53}, 89 (1999);
math.dg/9910002.

\bibitem{25} R. Gambini and J. Pullin, "Loops, Knots, Gauge Theories and
Quantum Gravity" (Cambridge Monographis in Mathematical Physics, 1996)

\bibitem{26} L. Smolin, "An Invitation to loop quantum gravity", published
in *Cincinnati 2003, Quantum theory and symmetries* 655-682; hep-th/0408048.

\bibitem{27} A. Corichi, J. Phys. Conf. Ser. \textbf{24}, 1 (2005);
gr-qc/0507038.

\bibitem{28} S. Hewson and M. Perry, Nucl.Phys.B \textbf{492}, 249 (1997),
hep-th/9612008.

\bibitem{29} M. J. Duff, Int. J. Mod. Phys. A \textbf{11, }5623\textbf{\ }%
(1996); hep-th/9608117.

\bibitem{30} L. Smolin, Phys. Rev. D \textbf{62}, 086001 (2000);\
hep-th/9903166.

\bibitem{31} L. Smolin, Nucl. Phys. B \textbf{739}, 169 (2006);
hep-th/0503140.

\bibitem{32} Y. Ling and L. Smolin, Phys. Rev. D \textbf{61}, 044008 (2000);
hep-th/9904016.

\bibitem{33} T. Thiemann, Class. Quant. Grav. \textbf{23}, 1923 (2006);\
hep-th/0401172.

\bibitem{34} A. Bj\"{o}rner, M. Las Vergnas, B. Sturmfels, N. White and G.
M. Ziegler, \textit{Oriented Martroids}, (Cambridge University Press,
Cambridge, 1993)

\bibitem{35} J. A. Nieto, Adv. Theor. Math. Phys. \textbf{8}, 177 (2004);
hep-th/0310071.

\bibitem{36} J. A. Nieto, Adv. Theor. Math. Phys. \textbf{10}, 747 (2006),
hep-th/0506106.

\bibitem{37} J. A. Nieto and M.C. Mar\'{\i}n, Int. J. Mod. Phys. A \textbf{18%
}, 5261 (2003); hep-th/0302193.

\bibitem{38} J. A. Nieto and M. C. Mar\'{\i}n, J. Math. Phys. \textbf{41,}
7997 (2000).

\bibitem{39} J. A. Nieto, J. Math. Phys. \textbf{45}, 285 (2004);
hep-th/0212100.

\bibitem{40} J. A. Nieto, "Maximal supersymmetry in eleven-dimensional
supergravity revisited and chirotopes"; hep-th/0603139.

\bibitem{41} A. Ashtekar, private communication, december 2006.

\bibitem{42} M. Montesinos, Class. Quant. Grav. \textbf{18}, 1847 (2001);\
gr-qc/0104068

\bibitem{43} J. F. Barbero, Phys. Rev. D \textbf{54}, 1492 (1996);
gr-qc/9605066.
\end{thebibliography}
\end{document}